# Evidence of tunable magnetic coupling in hydrogenated graphene

Shimin Cao[1], Chuanwu Cao[1], Shibing Tian[2], Jian-Hao Chen[1,3,4 †]

[1]*International Center of Quantum Material, School of Physics, Peking University, Beijing 100871, China*
[2] *Institute of Physics, Chinese Academy of Sciences, Beijing 100190, China*
[3]*Key Laboratory for the Physics and Chemistry of Nanodevices, Peking University, Beijing 100871, China*
[4]*Beijing Academy of Quantum Information Sciences, Beijing 100193, China*

**Abstract**

A lot of efforts have been devoted to understanding the origin and effects of magnetic moments induced in graphene with carbon atom vacancy, or light adatoms like hydrogen or fluorine. At the meantime, the large negative magnetoresistance (*MR*) widely observed in these systems is not well understood, nor had it been associated with the presence of magnetic moments. In this paper, we study the systematic evolution of the large negative *MR* of *in-situ* hydrogenated graphene in ultra-high vacuum (UHV) environment. We find for most combination of electron density ($n_e$) and hydrogen density ($n_H$), *MR* at different temperature can be scaled to $\alpha = \mu_B B / k_B (T - T^*)$, where $T^*$ is the Curie-Weiss temperature. The sign of $T^*$ indicates the existence of tunable ferromagnetic-like ($T^*>0$) and anti-ferromagnetic-like ($T^*<0$) coupling in hydrogenated graphene. However, the lack of hysteresis of *MR* or anomalous Hall effect below $|T^*|$ points to the fact that long-range magnetic order did not emerge, which we attribute to the competition of different magnetic orders and disordered arrangement of magnetic moments on graphene. We also find that localized impurity states introduced by H adatoms could modify the capacitance of hydrogenated graphene. This work provides a new way to extract information from large negative *MR* behavior and can be a key to help understanding interactions of magnetic moments in graphene.

## I. INTRODUCTION

Magnetism in graphene induced by non-magnetic defects has attracted a lot of attention in recent years[1-14]. From a theoretical point of view, point defects that remove $p_z$ orbitals in π-graphene can create single-electron occupying π states around the missing orbitals, therefore introduce local magnetic moments in graphene[1-4]. The moments can interact with each other ferromagnetically (antiferromagnetically) when they occupy the same (opposite) graphene lattice[2, 3]. Experimentally, such local magnetic moments can be directly imaged and manipulated[4, 5] using scanning tunneling microscopy. The moments are characteristic of two spin-split density-of-states peaks close to the Dirac point, with spatial extension of several nanometers[4, 5], suggesting the long-range exchange interaction between the moments can take place[4].

The existence of such local moments and interactions between them naturally leads to the question: what are the electrical transport properties of graphene with magnetic defects and whether there is global magnetic order in graphene with point defects? Since generating local moments on graphene surface in a controlled manner is challenging, magneto-transport of graphene with *in-situ* or *ex-situ* randomly generated carbon vacancy[6, 8] or adatoms like hydrogen[11, 12], fluorine[9, 10, 13] or ozone[7] has been studied. Spin-flip scattering[10] and Kondo effect[8] are found in graphene with point defects, which provides strong evidence that defects in graphene are indeed magnetic and the local moments can strongly couple with conduction electrons[8]. The magnetic moments are also found to scatter pure spin current and generate exchange field in graphene[14]. However, typical behavior of global magnetism like hysteresis loop of magneto-resistance (*MR*) or anomalous Hall effect is not reported. We notice that in the magneto-transport of graphene with local defects[8-11], colossal negative *MR* is widely observed yet not fully understood. These negative *MR* can extend to magnetic field as large as 9 T without saturation, and could be up to -95% [8-11]. The origin of this negative *MR* is not likely to be weak localization since typical feature of WL is within the range of B ≲ 0.5 T and within a magnitude on order $\Delta\sigma \sim e^2/h$ [8, 10, 15-17]. There are proposals that the negative *MR* is caused by strong localization or adatom-induced magnetism[9].

In this work, we report the magneto-transport of *in-situ* hydrogenated graphene in UHV environment. Similar non-saturating negative *MR* is observed. We find that for most combination of electron density ($n_e$) and H adatom density ($n_H$), the *MR* curves at different temperature can be scaled to one single trail when they are plotted against $\alpha = \mu_B B/k_B(T - T^*)$, where $T^*$ is the Curie-Weiss temperature. The magnitude and sign of $T^*$ can change systematically with $n_e$ and $n_H$, indicating tunable FM-like or AFM-like coupling. We find the relationship between $T^*$ and $n_H$ agrees with the theoretically predicted[18] power laws for AFM-like coupling at relatively low carrier density, which provides evidence for validity of the scaling method. At higher carrier density, FM-like coupling appears, instead of theoretically predicted non-magnetic state[19]. Like previous work, the evidence for long range global order is not observed, which may be due to competition of the FM-like and AFM-like magnetic orders or the disordered arrangement of the local defects in the system. We also find the quantum capacitance of hydrogenated graphene is substantially modified by localized impurity states.

## II. EXPERIMENTAL METHODS

The graphene sample used in this experiment is mechanically exfoliated on 300 nm/500 μm $SiO_2$/Si substrates using Scotch tape. Its thickness is decided with atomic force microscope (AFM) together with optical color code and subsequently confirmed by quantum Hall effect. Graphene device is fabricated using standard electron beam lithography (EBL) technique and metal contact of Cr 5 nm / Au 50 nm. The device is etched into Hall-bar geometry using reactive ion etching (RIE). Magneto-transport measurement is done in our home-built *in-situ* transport system, with ultra-high vacuum environment (up to $10^{-9}$ Torr) to reduce impurity absorption on sample surface between each round of hydrogen doping. The sample temperature of measurement is ranging from 5.8K to 62K, since we found higher temperature may cause desorption of H atoms from the graphene surface. The applied magnetic field is perpendicular to the graphene sample and the maximum field is 9 T. The graphene is doped using hydrogen atom source, which generates hydrogen atoms with radio-frequency (RF) plasma (13.56 MHz). The amount of hydrogen atoms is controlled by the RF power and doping time. The hydrogen plasma is usually generated at a vacuum of $10^{-2}$ torr and at RF power of 60W. Time of each round of doping varies from 5 s to 10 s, controlled with a mechanical shutter. Hydrogen ions are also generated in the process and are deflected with electrical field to avoid causing vacancy defects in graphene. Transport measurement is done using low-frequency lock-in amplifier technique. All the carrier density, unless specified, is calculated using parallel capacitor model $n_e = C_g(V_g - V_D)$, where $V_g$ is gate voltage, $V_D$ is Dirac point location, and $C_g$ for 300 nm SiO$_2$ is $7 \times 10^{10}\ cm^{-2}V^{-1}$.

### III. RESULTS AND DISCUSSION

*In-situ* experiments allow us to explore the effect of adatom doping on the transport properties of the sample in an almost continuous way. That means we can start from a pristine sample and give it a small amount of doping, which can be controlled by the doping power and doping time. Then we can stop and do thorough magneto-transport measurement after each round of doping. Usually the doping and measuring procedure will be repeated for 5 to 20 times, depending on the amount of hydrogen at each round of doping and on the final device quality. In the end, we obtain a curve of transport property versus the amount of doping. In this process, the sample is in UHV for most of the time (except during doping when the hydrogen partial pressure will increase to ~10$^{-2}$ torr), so it is hardly influenced by unwanted impurity absorption from the environment.

In our graphene *in-situ* hydrogenation experiment, the as-fabricated device has rather high quality, which is manifested by the well-developed quantum Hall effect, as shown in figure 1(a). After each round of hydrogenation, the sample still shows typical V-shape conductivity vs. $V_g$ curves [figure 1(b)], with the mobility reduced [figure 1(c)], as expected for increasing disorder in system. The Dirac point of graphene shifts to negative [figure 1(c)], which means hydrogen atoms bring extra electron doping to graphene, consistent with previous APRES study [20]. We notice that $\sigma_{xx}$ is proportional to carrier density when Fermi surface is far from the Dirac point [figure 1(b)], which is a typical result of the charged impurity scattering [21] and defect scattering [6]. For graphene with screened charged impurity [21] or mid-gap states [6], we have $\sigma_{xx} = C(e^2/h)(|n|/n_{imp})$, where $C$ is a constant depending on the details of impurity, $n$ is carrier density in graphene and $n_{imp}$ is impurity density. Hydrogen atoms chemisorbed on graphene essentially change the original sp$^2$ hybridization of carbon atom to sp$^3$ and remove the p$_z$ orbital [4, 22]. In this sense, H chemisorption is equivalent to creation of carbon vacancy on graphene. For the latter case, previous work[6] gives the value of $C$ to be about 4. Here, we use the same value as an estimate for the order of magnitude. Figure 1(d) shows the estimated hydrogen density on graphene. Maximum hydrogen density is of order 10$^{13}$ cm$^{-2}$, which gives a maximum doping coverage of 0.28% and an average H atom distance down to 2.1nm.

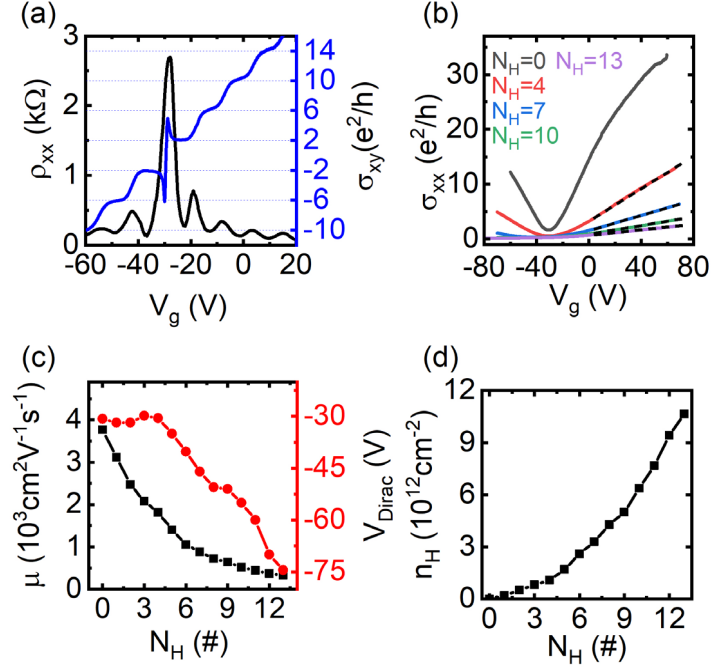

FIG 1. Transport property of pristine and hydrogenated graphene. (a) Quantum Hall effect of pristine graphene at $T$ = 5.8 K, $B$ = 9 T. (b) Conductivity of graphene after hydrogenation at $T$ = 5.8 K. $N_H$ = 0, 1, 2,… means hydrogenation round, $N_H$ = 0 corresponds to pristine graphene. Dashed lines are straight lines as guide for the eye. (c) Mobility at $T$ = 5.8 K and Dirac point shift of hydrogenated graphene. (d) Estimated hydrogen adatom density on graphene after each round of hydrogenation.

Since the Dirac point is far from zero after several rounds of H doping, we focus on the magneto transport on the electron side. Figure 2 shows the *MR* of hydrogenated graphene at different combinations of electron density ($n_e$) and hydrogen density ($n_H$). We find for most case the *MR* is negative. The negative *MR* is not saturating up to 9 T, and the maximum value can be -52%. The negative *MR* is found to decay with rising temperature. Similar negative *MR* has already been observed in graphene with carbon vacancy [8] and fluorine adatoms [9, 10].

The origin of such *MR* is multiple. For small field part ($B$ < 0.5 T), the *MR* is confirmed to originate from weak localization [8] or weak anti-localization, which we will discuss elsewhere. Here we focus on the large field *MR*. At the first five rounds (see figure S2) of hydrogenation, Shubnikov de Haas (SdH) oscillations clearly exist, contributing to some *MR* features. For negative *MR* not affected by SdH oscillations (e.g., with higher impurity density and lower carrier mobility), there are proposals [9] that similar phenomenon can be caused by strong localization[23, 24] or adatom-induced magnetism [10]. Strong localization model suggests the graphene is divided into disconnected conducting areas, and charge carriers can hop from one area to another. In this regime, the graphene resistivity should be described by variable range hopping (VRH) model [23, 25]. We analyzed the relationship of resistance and temperature of our hydrogenated graphene, and found that it does not follow either ES-type or Mott-type VRH model (figure S3) [23, 25]. In fact the localization is relatively weak even for the high H density samples (figure S3b&3d). Therefore, strong localization can be ruled out for our experiment and the negative *MR* can only be caused by the magnetic moments on graphene

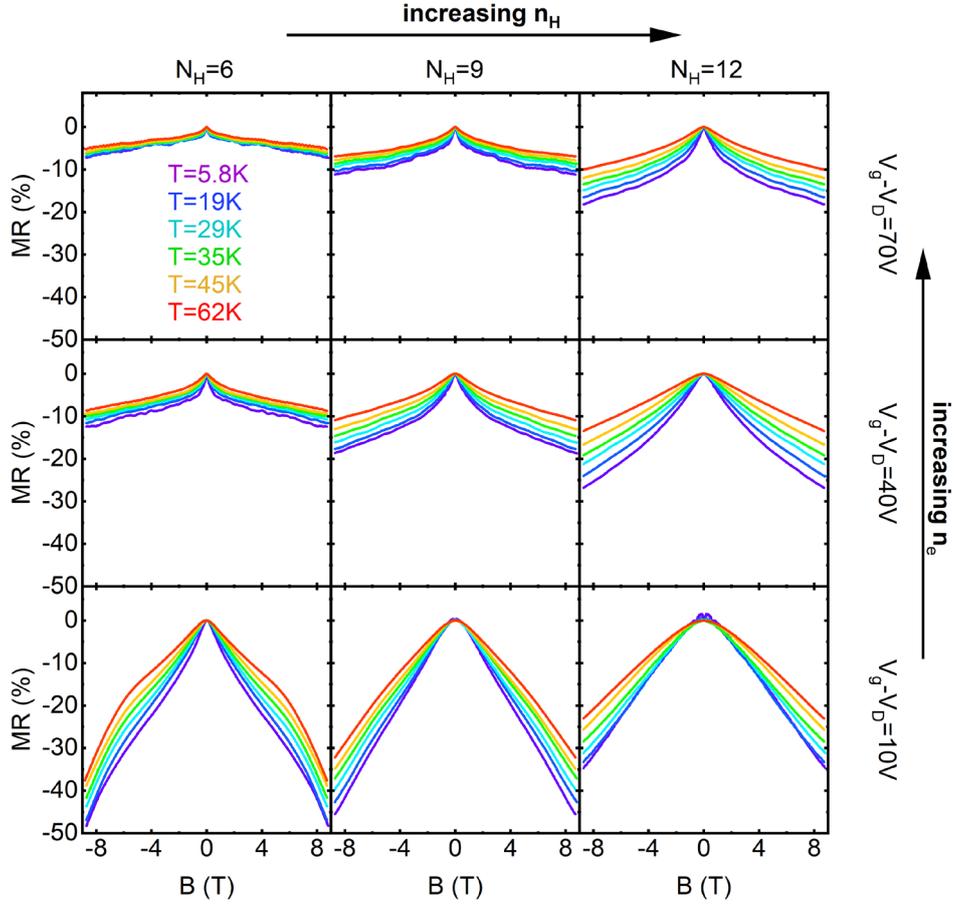

FIG 2. Magnetoresistance of hydrogenated graphene at different hydrogen density ($n_H$) and electron density ($n_e$). All plots use the same scale for comparison. $N_H$ means hydrogenation round. Different color stands for different temperature.

Such negative *MR* in magnetic materials is rather normal, like diluted magnetic semiconductors [26-29], alloys [30-32], as well as oxides [33, 34]. Negative *MR* in magnetic materials usually origins from charge carriers scattered by the isolated magnetic moments, which is a function of $MR \equiv MR(M, \alpha)$ [26, 34], where *M* is the total magnetization of the material and $\alpha \equiv \mu_B B / k_B (T - T^*)$. Here $T^*$ is the Curie-Weiss temperature, describing the exchange effect between the magnetic moments. If the material is paramagnetic or above its critical temperature, we can write the low-field magnetization as $M = \chi H \propto H/(T - T^*) = M(\alpha)$. Thus we can express *MR* as a function of *α*, e.g., $MR = MR(\alpha)$ before magnetic ordering occurs.

Important information can be readily extracted from the fact that $MR = MR(\alpha)$, without going into the detailed functional form of *MR*, which could depend on material-specific parameters. Since there is no explicit dependence on *T* in $MR(\alpha)$ curves, for a certain combination of $n_e$ and $n_H$, the form of *MR* should be invariant at different temperatures when plotted against $\alpha \equiv \mu_B B / k_B (T - T^*)$. In another word, *MR* for any specific $n_e$ and $n_H$ at all *T* above magnetic ordering temperature can be scaled into one single trail when we plot *MR* versus *α*, if we used the right $T^*$. For example, in figure 3(a), we plot *MR* curves versus *α* for three different values of $T^*$ for the 12$^{th}$ hydrogenation round and at electron density $V_g$-$V_D$ = 40 V. Only data corresponding to temperature higher than 25K and magnetic field lower than 3T [13] is used to fulfill the fore mentioned criteria. As we can see, the curves can perfectly collapse at $T^*$ = -7.2K. Mathematically, $T^*$ is obtained by minimizing the square sum $S = \sum(MR_i - \overline{MR})^2$, detailed description can be found in supplementary information (Figure S4). Repeating that procedure for other $n_e$ and $n_H$ gives a two-dimensional plot of $T^*$ as shown in figure 3(b), which reveals the phase diagram of the effective coupling strength for the magnetic moments in hydrogenated graphene. Since the sign of the Curie-Weiss temperature indicates the type of magnetic coupling. Thus we can clearly find that the phase diagram in figure 3(b) is divided into ferromagnetic-like ($T^*$>0) and antiferromagnetic-like ($T^*$<0) regime.

Previous theoretical work [18, 19, 35] predicted that AFM order can exist in dilute doped graphene at low carrier density, if the magnetic moments are evenly distributed on both sublattices. In this case, |$T^*$| is predicted to be proportional to $n_H^{\gamma/2}$, where $\gamma$ is the decaying index of RKKY interaction $J(r) \propto 1/r^\gamma$. In graphene, the value of $\gamma$ is found to be 3, due to the absence of extended Fermi surface[18, 19]. We take |$T^*$| under various electron densities and plot them against $n_H$ in log-log scale, as shown in figure 3(c). We find $|T^*| \propto n_H^{3/2}$ can perfectly fit our scaling data points, which proves the validity of our scaling method.

At higher carrier density, the RKKY interaction will develop Friedel oscillation [19, 36]. It gives a random sign to the exchange interaction of magnetic moments with distance larger than $r > 1/\sqrt{n_e}$. So ordered magnetic phase is expected to be suppressed at a sufficiently large $n_e/n_H$ ratio[19]. In our phase diagram, the boundary of AFM order is indeed close to a line corresponding to $n_e/n_H \simeq 3.1$, as the dashed line in figure 3(b). However, over this boundary, $T^* > 0$, indicating that FM-like magnetic coupling appears, different from the predicted non-magnetic state[19] or competition between AFM and Kondo effect [37]. We also plot $T^*$ vs. $n_H$ in this regime in figure 3(d) using log-log scale, and surprisingly find the relationship $T^* \propto n_H^{-3/2}$ can describe our scaling data quite well. We notice this fits the theoretical prediction for the case that magnetic moments are only on one set of graphene sublattices [38]. Even though our experimental condition is definitely different from the premise of ref. [38], the surprising agreement might indicate the emergence of FM-like coupling and it might point to something deeper than a mere coincidence. Clearly, more theoretical work is required to explain our experimental observation.

Another point to note is the relation between magnetic ordering temperature $T_N$ ($T_c$) and the Curie-Weiss temperature $T^*$. In the AFM-like regime, the |$T^*$| can be higher than 80 K; in the FM-like regime, $T^*$ can reach about 20 K. But no evidence of long-range FM order, such as hysteresis loop in $MR$ and $R_{xy}$ is observed when $T < T^*$ in the FM-like regime. So the actual ordering temperature $T_N$ ($T_c$) must be much lower than the Curie-Weiss temperature. A possible explanation could be that the long-ranged order is suppressed by the competition between FM and AFM order; another possibility arises from the disordered location of the magnetic moments in the system. Indeed, in spin liquid or systems with magnetic frustration [39, 40], the actual critical (ordering) temperature can be 5 to 10 times smaller than the Curie-Weiss temperature. Future work at lower temperature will help to clarify this issue.

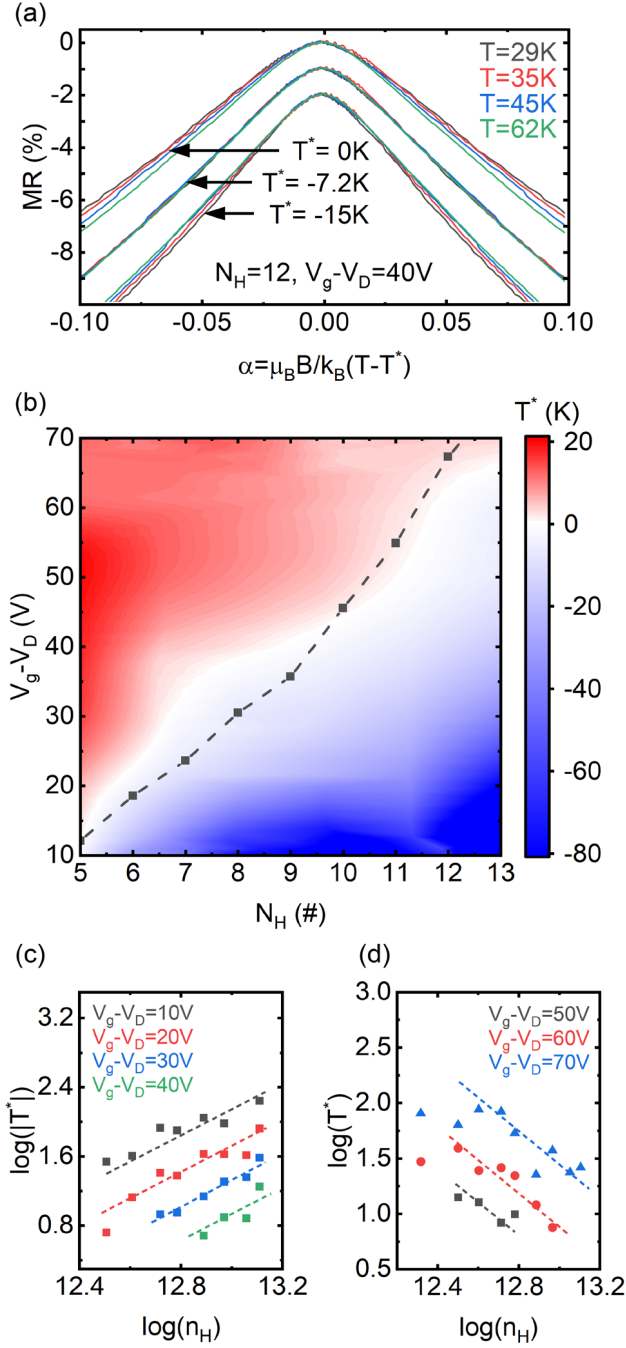

FIG 3. Scaling and magnetic phase diagram of hydrogenated graphene. (a) Scaling of *MR* for $N_H$ = 12, $V_g$-$V_D$ = 40 V. $N_H$ means hydrogenation round. (b) Phase diagram of scaled Curie-Weiss temperature $T^*$. Dashed line corresponds to $n_e/n_H \simeq 3.1$. (c) $|T^*|$ vs. hydrogen density $n_H$ in the AFM-like regime in (b), in log-log scale. Dashed lines are $|T^*| \propto n_H^{3/2}$, as guide for the eye. (d) $T^*$ vs. $n_H$ in the FM-like regime in (b), in log-log scale. Dashed lines are $T^* \propto n_H^{-3/2}$, as guide for the eye.

Finally, we report our finding on the Hall density of states (DOS) of hydrogenated graphene. Figure 4(a) shows the Hall resistance of pristine and hydrogenated graphene. There is no sign of anomalous Hall effect after hydrogenation, but the slope of the Hall resistance is modified. Figure 4(b) shows the measured Hall carrier density and linear fitting of $n = C_H(V_g - V_D)$, where $C_H$ stands for the capacitance calculated from Hall effect carrier density. The left axis of figure 4(c) shows the fitted capacitance $C_H$ as a function of the hydrogenation round $N_H$ and at different temperatures from 5.8K to 62K. Before hydrogenation, $C_H$ is close to the value of 300 nm $SiO_2$ dielectric parallel capacitor [41]. We find that $C_H$ drops almost 50% after hydrogenation and it is almost temperature independent.

The origin of the modification of $C_H$ can be understood as the following: First of all, hydrogen atoms would do very little in changing the dielectric constant of the 300nm $SiO_2$ substrate, especially for the portion of the $SiO_2$ under the protection of the graphene flake. Thus it is not a changing dielectric constant that is responsible for the drop in $C_H$. Second, the change in Hall resistance slopes cannot be related to magnetic ordering in the sample, otherwise, the Hall signal should depend on temperature prominently, similar to behavior of the longitudinal *MR*. One possible reason is the localized impurity states due to hydrogenation. Previous work pointed out that there will be impurity states around the Fermi level for graphene with vacancy defects [42] or H adatom impurity [43]. And the impurity states are found to be quasi-localized [42-44], such that a moderate change in temperature (from 5.8K to 62K) will not cause delocalization[45], and the only way to deplete such localized electron would be by the change of Fermi level via $V_g$. These localized states will reduce the carrier density that can be observed in transport, leading to an reduction in $C_H$. The physical picture of such localized states is illustrated in figure 4(d). Furthermore, if we plot the shift of the Dirac point together with change in $C_H$, as shown in the right axis of figure 4(c), we find the two curves almost collapse. Since one hydrogen adatom provides a fixed doping to the sample, therefore the amount of impurity states created by hydrogen absorption should be reflected by a proportional shift in the Dirac point in graphene.

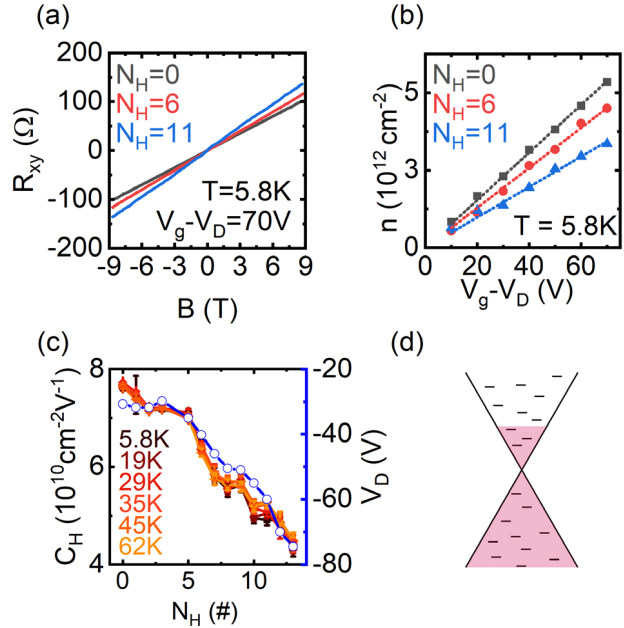

FIG 4. Density of states in hydrogenated graphene. (a) Hall resistance of pristine and hydrogenated graphene. (b) Dependence of Hall carrier density on gate voltage. Dashed lines are linear fitting to $n = C_H(V_g - V_D)$. (c) Solid dots: capacitance at various temperatures calculated from Hall carrier density. Open circle: Dirac point position. (d) Illustration of hydrogenated graphene band structure. Short dash stands for energy levels of localized impurity states.

## IV. CONCLUSION

In summary, we studied the magneto-transport of hydrogenated graphene with *in-situ* doping method. We find that hydrogenated graphene shows large negative *MR*, similar to graphene with carbon vacancy [8] and fluorine adatoms [9, 10]. By scaling the *MR* curves using proper Curie-Weiss temperatures, two different regimes are found with FM-like and AFM-like magnetic coupling, respectively. The type of magnetic coupling is jointly tunable by carrier density and hydrogen density, and there is no evidence of long-ranged magnetic ordering. The existence of AFM-like coupling is a result of RKKY interaction of H adatom induced magnetic moments on graphene. But the appearance of FM-like coupling is unexpected, which requires more theoretical investigation. Finally, we found that hydrogen adatom induced localized impurity states has a substantial impact in the charge carrier density of states. This work provides a novel method to extract key information from large negative *MR* in similar physical systems and could be a key factor to better understand 2D magnetism.

## V. ACKNOWLEDGEMENT

The authors thank X. C. Xie, S. Q. Shen and X. D. Xu for helpful discussion and C. Y. Cai for help on experiment. This project has been supported by the National Basic Research Program of China (Grant Nos. 2019YFA0308402，2018YFA0305604), the National Natural Science Foundation of China (NSFC Grant Nos. 11934001，11774010，11921005).

†Corresponding author: chenjianhao@pku.edu.cn

# Supplementary information for

# Evidence of tunable magnetic coupling in hydrogenated graphene

Shimin Cao[1], Chuanwu Cao[1], Shibing Tian[2], Jian-Hao Chen[1,3,4 †]

[1]International Center of Quantum Material, School of Physics, Peking University, Beijing 100871, China
[2] Institute of Physics, Chinese Academy of Sciences, Beijing 100190, China
[3]Key Laboratory for the Physics and Chemistry of Nanodevices, Peking University, Beijing 100871, China
[4]Beijing Academy of Quantum Information Sciences, Beijing 100193, China

Content in the supplementary information includes:
1. Schematic illustration of the *in-situ* experiment setup
2. Shubnikov de Haas (SdH) oscillation for different hydrogenation runs
3. Temperature dependent resistance of hydrogenated graphene at zero magnetic field
4. Least square sum (LSS) calculation to determine the Curie-Weiss temperature $T^*$

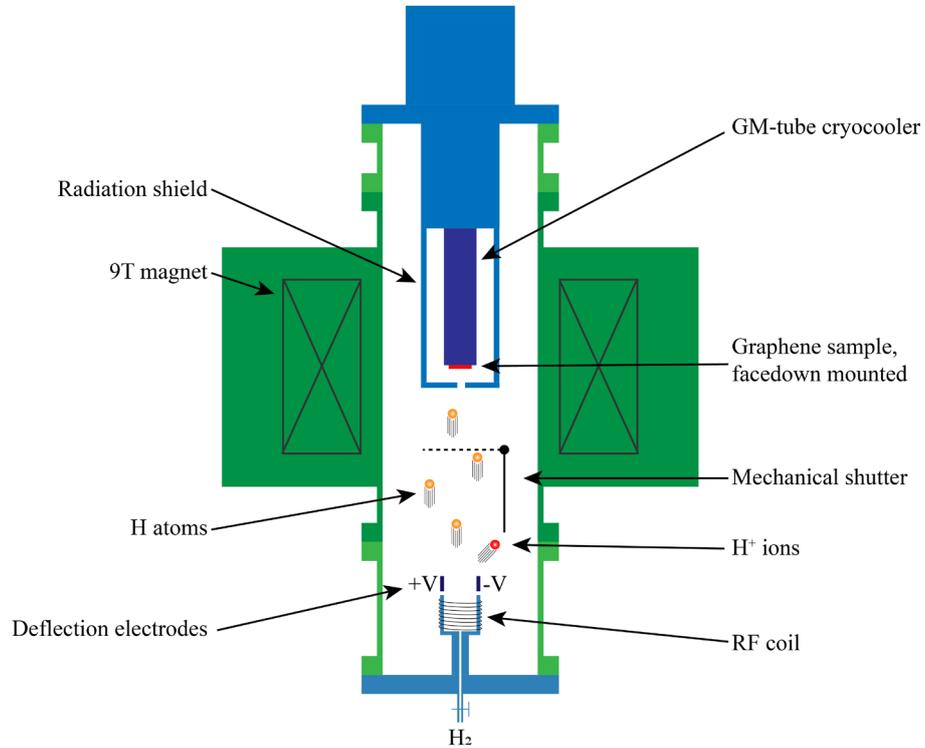

Figure S1. Schematic illustration of the *in-situ* experiment setup. Base temperature of the sample is 5.8K.

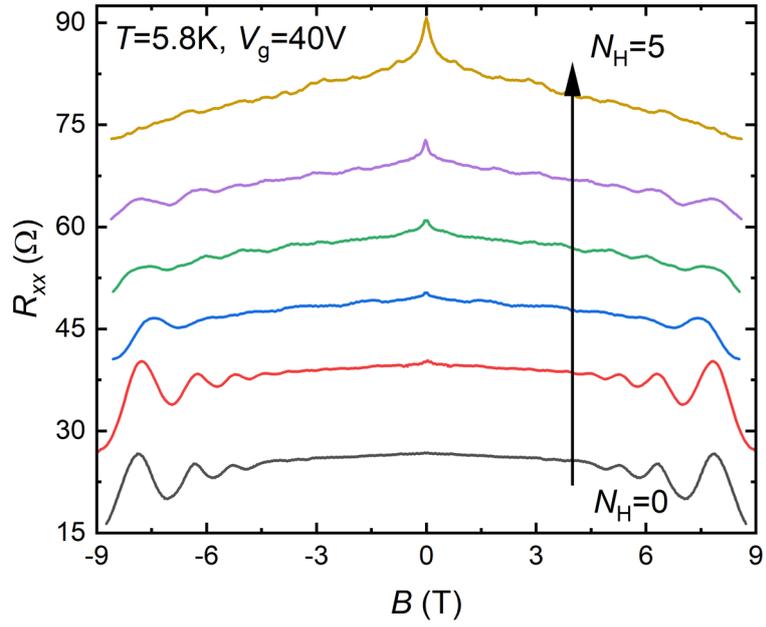

Figure S2. Shubnikov de Haas (SdH) oscillation for different hydrogenation runs. At 5$^{th}$ round of hydrogenation, the SdH oscillation essentially disappears.

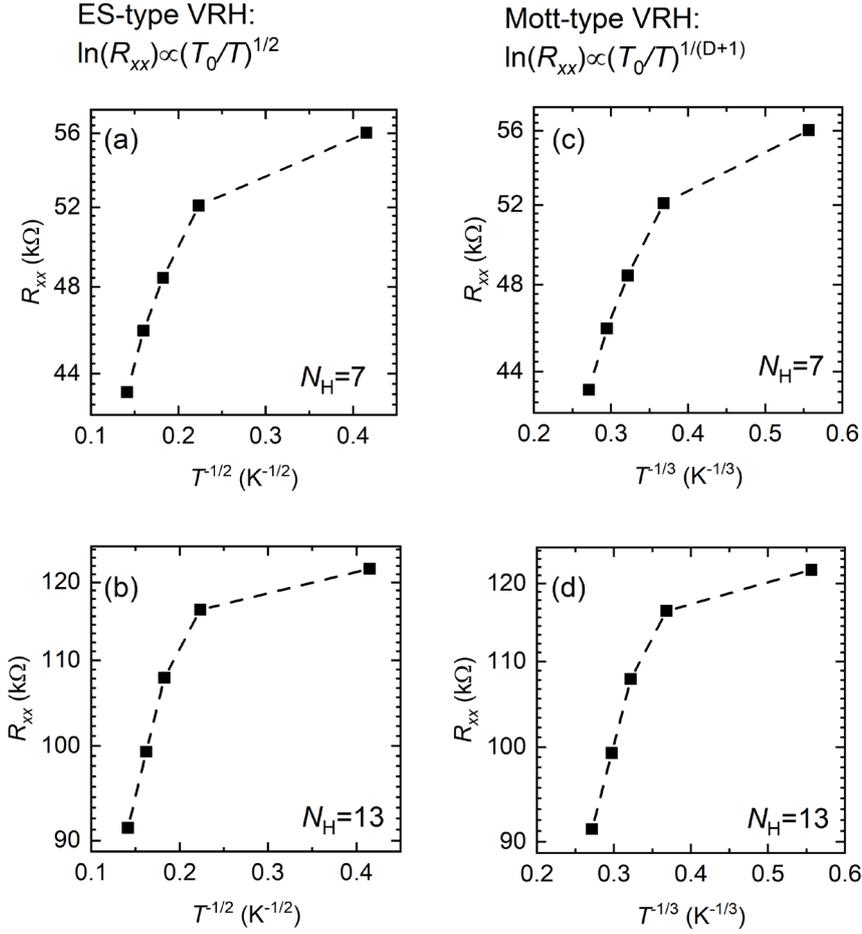

Figure S3. Temperature dependent resistance of hydrogenated graphene at zero magnetic field. If the transport can be described by ES-type VRH model, $\ln(R_{xx})$ should be proportional to $T^{-1/2}$, which should give a straight line in (a) or (b). Similarly, if the transport can be described by Mott-type VRH, one should see a straight line in (c) or (d). Here the dimension D is equal to 2 for graphene for the formula at the top of the figure S3(c). Thus the transport behavior of hydrogenated graphene at zero magnetic field cannot be described by the VRH model, and it is far away from the strong localization regime.

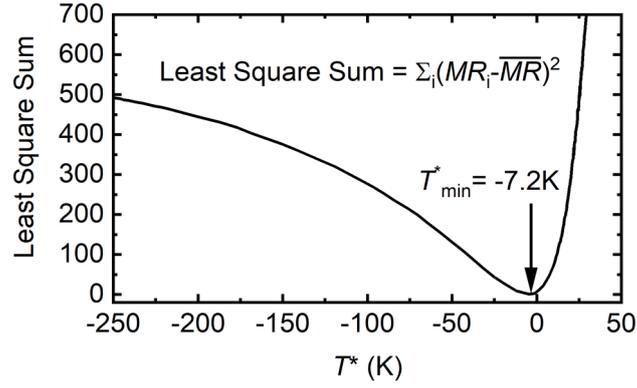

Figure S4. Least square sum (LSS) calculation to determine the Curie-Weiss temperature $T^*$. Data is from $N_H = 12$, $V_g$-$V_D = 40V$, as shown in figure 3(a). $MR_i$ stand for $MR$ curves at different temperatures for a particular carrier density $n$ and hydrogen density $n_H$, $\overline{MR}$ is average of all the $MR_i$. A "total distance" is defined by the LSS $\Sigma_i(MR_i - \overline{MR})^2$. We calculate the LSS for a range of trail Curie-Weiss temperatures $T^*_{trail}$, and we get a smooth curve with only one global minimal point. The $T^*_{trail}$ that minimizes LSS is then selected to be Curie-Weiss temperature $T^*$ for that particular carrier density $n$ and hydrogen density $n_H$. For all other combination of carrier density and hydrogen density, the LSS curves have similar shape with this one.